\documentclass[reprint,amsmath,amssymb,aps]{revtex4-2}

\usepackage{graphicx}
\usepackage{dcolumn}
\usepackage{bm}

\begin{document}

\preprint{AIP/123-QED}

\title{Dynamics of order-disorder and complexity for interacting bosons in optical lattice}
\author{Rhombik Roy$^1$, Barnali Chakrabarti$^1$, N. D. Chavda$^2$, M. L. Lekala$^3$}
\affiliation{$^1$Department of Physics, Presidency University, 86/1   College Street, Kolkata 700073, India. \\ $^2$Department of Applied Physics, Faculty of Technology and Engineering, The Maharaja Sayajirao University of Baroda, Vadodara-390001, India. \\  $^3$Department of Physics, University of South Africa, P.O. Box 392, Pretoria 0003, South Africa}

\date{\today}

\begin{abstract}
The present work reports on the dynamical measures of order, disorder and complexity for the interacting bosons in optical lattice. We report results both for the relaxed state as well as quench dynamics. Our key observations are: (1) Lattice depth can be taken as order-disorder parameter. (2) The superfluid to Mott insulator transition can be treated as `order-disorder' transition. Our main motivation is to find how the system organize by itself during quench and how it optimizes the complexity. We find dynamical measures of order and disorder are more sensitive tool than entropy measures. We specifically calculate the time scale of entry and exit of different phases during time evolution. Initially the system exhibits collapse revival trend, however gradually looses its ability to turn back to superfluid phase and finally Settle to Mott insulator phase.
\end{abstract}

\maketitle

\section{Introduction} \label{intro}
In the recent years, different measures of complexity have been proposed in several scientific disciplines~\cite{complexity6,complexity7,complexity8,complexity9,complexity10,complexity11,complexity12,complexity13}. Out of these, information theoretical measures and entropic uncertainty relations in quantum mechanical systems become the trademark~\cite{ent1,ent2,ent3,ent4,ent5}. In this context, the concept of complexity has also attracted considerable interest~\cite{complexity5,complexity14,complexity15,complexity16,complexity17,complexity18,complexity19,complexity20}. Complexity means how a physical or biological system will organize themselves in response to some change in the external parameters. It is usually believed that with increase in number of particles $N$, the complexity would also increase. Although there are several definitions of complexity in the literature ~\cite{complexity1,complexity2,complexity3,complexity4}, the usual definition of complexity $\Gamma_{\alpha \beta}$ was introduced by Shiner, Davison and Landsberg~\cite{complexity_PhysRevE} known as SDL complexity. The other statistical measures of complexity $\Bar{C}$ was defined by L\'{o}pez- Ruiz~\cite{LOPEZRUIZ} and known as LMC measure. In the context of measuring Shannon information entropy, complexity is taken as the most efficient measure of order and disorder exist in a system~\cite{complexity21,complexity22,complexity23,complexity24,complexity25,complexity26,complexity27}. The most common case is the `convex' type complexity where it is minimum both for completely ordered and disordered system. Some system also exhibit complexity which are either increasing or decreasing function of disorder~\cite{complexity_PhysRevE}. The concept of statistical complexity was first successfully applied in atomic system by the group of Panos~\cite{panos}. Both SDL and LMC were studied as a function of atomic number $Z$ and main interest was to explore the connection of the periodicity of shell structure with the complexity measures.\\
In the present work, we are interested in the SDL complexity measures for interacting bosons in the optical lattice. The realization of fully controlled quantum many-body system is an outstanding challenge since past years. Interacting bosons in an external trap at ultracold temperature allow an unprecedented experimental control. Interacting bosons in optical lattice feature a variety of quantum phases--- superfluid phase (SF), Mott insulator phase (MI), fragmented Mott insulator (FMI)~\cite{rhombik_pra}. In the pioneering experiment of Greiner et. al.~\cite{greiner1}, a quantum phase transition is observed in a Bose Einstein condensate (BEC) kept in an optical lattice potential with repulsive inter atomic interaction.  It is observed that weakly interacting bosons in shallow optical lattice exhibits a superfluid phase. In the SF phase, the atoms exhibit long range phase coherence across the lattice. Increasing the depth of the lattice, SF phase makes a transition to Mott insulator phase. In MI phase, atoms are localized in the individual lattice sites and phase coherence across the lattice is lost. The quantum phases are studied by Bose-Hubbard model~\cite{bosehubbard1,bosehubbard2}, {\textit{ab-inito}} many-body technique---multiconfigurational time dependent Hartree for bosons (MCTDHB)~\cite{budha-order-parameter,budha2,rui-lin1,rhombik_pra}. The many-body features are characterised by distinct measures of many-body correlation, collapse and revival dynamics in lattice depth quench~\cite{rhombik_jpb}. The collapse-revival dynamics in the measure of correlation shows the exact behaviour of Fig.2 of ref ~\cite{greiner2}.\\
Our present work is focused to characterize the quantum phases of the interacting bosons in optical lattice through the measures of order, disorder and complexity. Although the measure of complexity uses the defined notations of order and disorder, but it exhibits considerable degree of flexibility. In the SF to MI transition, the lattice depth is experimentally controlled and theoretically corresponding dynamics is recently studied~\cite{budha-order-parameter,budha2,rhombik_jpb,rhombik_pra}. Our present calculation exhibits how the system optimizes the complexity. Of course we need to resolve whether the lattice depth parameter can be taken as a parameter of disorder. If so, it leads another open question whether SF to MI transition can be considered as an order-disorder transition. The relevant issue is to study how the system becomes self organized during the sudden quench. \\
We consider a system of $N=3$ bosons interacting with a contact interaction $W(x_i-x_j)= \lambda \sum_{i<j}^{N} \delta(x_i-x_j) $ in one dimensional three well optical lattice, $\lambda$ is the strength of interaction and the bosons are trapped in a lattice of form $V_{OL}(x)= V_0 sin^2 (kx)$, where $V_0$ is the depth of the optical lattice and $k$ is the periodicity of the lattice. Lattice depth is experimentally tunable. In the present work, lattice depth is tuned to achieve SF phase to MI phase transition. The Bose Hubbard model is widely successful to describe the phase transition from SF to MI state. However the most challenging many-body physics is expected in the regime of strongly interacting boons in shallow optical lattice ---- it is beyond the scope of Bose Hubbard physics. We solve the many-body  Schr\"odinger equation at a high level of accuracy by MCTDHB method (presented in Sec.~\ref{method}).\\
Following the definition of information entropy measures, order($\Omega$), disorder($\Delta$) as given in Sec.\ref{qnt}, we calculate complexity $\Gamma_{\alpha\beta}= \Delta^{\alpha} \Omega^{\beta}$. We observe that lattice depth parameter acts as a measure of order-disorder and complexity shows all the three types behaviour as defined in literature~\cite{complexity_PhysRevE} depending on the choice of $\alpha$ and $\beta$. We also note that SF phase is characterised with maximum order and MI phase exhibits maximum disorder. The `SF-MI' transition can be termed as `order-disorder' transition in the language of statistical measures.
For lattice depth quench, we prepare an initial state which is a pure SF phase. Instantaneous increase in the lattice depth triggers the system to go to MI phase. From the time evolution of entropy measures, we further calculate the time dynamics of order, disorder and complexity.  We observe collapse revival dynamics in all measures in short time dynamics as discussed in details in Sec.\ref{result}. We again observe that maximum order is associated with SF phase and MI phase is associated with maximum disorder.
Of course, our calculation for time dynamics reconfirm that SF to MI phase can be termed as `order-disorder' transition. However, the intriguing observation is that we  are able to focus on the time scale of entry and exit of different phases over several cycles. We also able to find out the holding time of Mott phase which is exposed as a plateau region in the time dynamics of order and disorder. We clearly demonstrate that long time dynamics of order-disorder is more sensitive tool as we are able to calculate the time scale of dynamical evolution. To showcase how the system is able or disable to organize the impact of the quench, we measure several different kinds of complexity. $\Gamma_{1,1}$ is the most fundamental which keep equal weight in order and disorder. $\Gamma_{0,4}$ is the same with zero disorder and $\Gamma_{1/4,0}$ is with zero order. The gross observation in the time evolution of all the three measures in complexity, is same ---- collapse and revival in the short time scale. The system of bosons can adjust the effect of quench and is able to make a transition to MI phase with a subsequent revival to SF phase. However in the long time dynamics, the system gradually looses its ability to turn back to SF phase and finally settle to MI phase. 

The paper is structured as follows. In Sec.\ref{method}, we introduce the setup and the necessary theory. Sec.\ref{qnt}  deals with the basic equation to measure the different quantities. Sec.\ref{result} explains our numerical results and explanations. Sec.\ref{conc} draws our conclusion.
\begin{figure}
    \centering
    \includegraphics[scale=0.4, angle=-90]{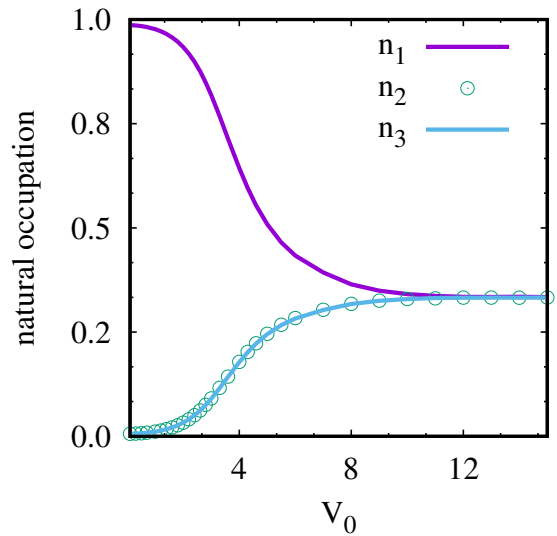}
    \caption{Population of the first three natural orbitals as a function of lattice depth ($V_0$). As $V_0$ increases, the occupation in the first orbital starts to decrease while another two orbitals start to contribute. At $V_0 =10.0$, the state becomes threefold fragmented ($ n_1 \simeq n_2 \simeq n_3 \simeq \frac{1}{3}$).  }
    \label{fig1}
\end{figure}

\section{Setup and methodology} \label{method}
In the present work, we consider $N=3$ bosons confined in one-dimensional optical lattice and interacting with contact inter-particle interaction. This one dimensional regime is easily achieved by tight transverse confinement. Quantum many-body effect is also important in such reduced dimension as the quantum fluctuation plays an important role. 
The Hamiltonian for $N$ interacting bosons in $1D$ optical lattice is given by
\begin{equation} \label{hamiltonian}
    H= \sum_{i=1}^{N}\left( -\frac{1}{2} \frac{\partial^2}{\partial x^2} + V_{OL}(x_i)  \right) + \sum_{i<j} \hat{W}(x_i-x_j)
\end{equation}
Where $V_{OL}$ represents the external lattice potential and 
$\hat{W}(x_i-x_j)$ is the two-body interaction.  We make the Hamiltonian dimensionless by dividing it by the factor $\frac{\hbar^2 }{mL^2}$, where $m$ is the mass of the bosons and  $L$ is some arbitrary length scale. The Hamiltonian is scaled in terms of recoil energy, $E_R = \frac{\hbar^2 k^2}{2m}$.  Thus the time is expressed in unit of $\frac{\hbar}{E_R}$ and unit of distance becomes $k^{-1}$. We use natural units ,i.e., $\hbar= m=k=1$. 
We fix up the grid at $x_{min}= -\frac{3 \pi}{2}$ to $x_{max}= \frac{3 \pi}{2}$ such that we consider three wells. For $N=3$ bosons, we choose $V_0$ and $\lambda$ such that it allowes superfluidity in the initial state. We find the stationary solution of the many-body Schr\"odinger equation by MCTDHB implemented in the MCTDH-X software\cite{mctdhb1,mctdhb2,mctdhb3} with periodic boundary condition.
In MCTDHB, the wave-function of the interacting bosons is expanded over a set of permanents which are the symmetrized bosonic states of $N$ bosons distributed over $M$ single particle states.
\begin{equation}
   \vert \Psi(t)\rangle = \sum_{\bar{n}}^{} C_{\bar{n}}(t)\vert \bar{n};t\rangle,
\label{many_body_wf}
\end{equation}
The vector $\Vec{n} = (n_1,n_2, \dots ,n_M)$ represent the occupation of the orbitals and $n_1 + n_2 + \dots n_M = N$ preserve the total number of particle.
\begin{equation}
 \vert \bar{n};t\rangle  = \prod_{i=1}^{M}
\left( \frac{ \left( b_{i}^{\dagger}(t) \right)^{n_{i}} } {\sqrt{n_{i}!}} \right) \vert vac \rangle.
\label{many_body_wf_2}
\end{equation}
$b_k^\dagger (t)$ creates a boson occupying the time-dependent orbital $\phi_k (x,t)$. The number of possible configuration is $ \left(\begin{array}{c} N+M-1 \\ N \end{array}\right)$. It is important to note that both the expansion coefficients ($C_{\Vec{n}} (t)$) and time dependent orbitals ($\phi_i (x,t)$) that build the permanents $\vert \Vec{n}, t \rangle$ are time dependent and fully variationally optimized quantities. Thus MCTDHB has been established as the most efficient way to solve the time dependent many-body Schr\"odinger equation~\cite{mctdhb_exact1,mctdbh_exact2,mctdhb_exact3}. The efficiency of MCTDHB is to make the sampled Hilbert space dynamically follow the time evolution of the many-body system. MCTDHB has been widely used in different theoretical calculations~\cite{rhombik_pra,rhombik_quantumreports,rhombik_aipconference,rhombik_epjd} and results are very close to experimental predictions~\cite{mctdhb_exp1,mctdhb_exp2}. 
For $M \rightarrow \infty$ limit, as the set of permanents $\vert \Vec{n;t} \rangle$ span the complete Hilbert space, the expansion is exact. But during computation, we limit the size of the Hilbert space. As the permanents are now time dependent, a given degree of accuracy is achieved with the truncated basis as compared to time-independent basis. It is also proved that significant computational advantage is achieved over exact diagonalization~\cite{barnali_axel}. To solve the time dependent many-body Schr\"odinger equation $\hat{H} \vert \psi \rangle = i \frac{\partial \vert \psi \rangle}{\partial t}$ for the wave function $\vert \psi \rangle$, we calculate the time evolution of the coefficients $C_{\Vec{n}} (t)$ and the orbitals $\phi_i (x,t)$. We utilize variational principle~\cite{variational1,variational2,variational3,variational4} to obtain the equation of motion of the time dependent coefficient and orbital~\cite{mctdbh_exact2,td1,mctdhb_exact1,td2,td3}. Finally the coupled nonlinear integrodifferential equations are solved by MCTDHB package~\cite{mctdhb3}.
For the calculation of the eigenstates of the Hamiltonian, we prepare the MCTDHB equations in imaginary time --- called improved relaxation method. For quench dynamics, we consider the total Hamiltonian
\begin{equation}
\hat{H}(x_1,x_2, \dots x_N)= \sum_{i=1}^{N} \hat{h}(x_i) + \Theta(t) \sum_{i<j=1}^{N}\hat{W}(x_i - x_j) 
    \label{propagation_eq}
\end{equation}
where $\hat{h}(x)$ is the one-body part includes the external trap and kinetic energy. $\Theta (t)$ is the Heaviside step function of time $t$ which trigger the quench at $t=0$. 

\begin{figure}
    \centering
    \includegraphics[scale=0.18, angle=-90]{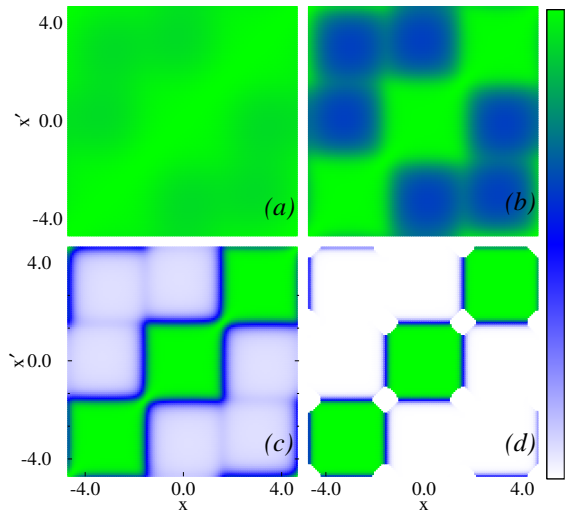}
     \caption{First-order normalized correlation function $g^{(1)}(x_1^{\prime},x_1,t)$ as a function of lattice depth ($V_0$). (a) $V_0= 0.7$ is in superfluid phase. inter-well coherence as well as intra-well coherence is maintained throughout the lattice. (b)$V_0= 2.8$; It is the initiation point of MI state. Diagonal correlation starts to build up which means the loss of inter-well coherence. (c) $V_0= 5.0 $; MI is the dominating phase. Intra-well coherence is prominent whereas inter-well coherence faded away. (d) $V_0= 10.0$ corresponds to pure MI phase where we find absolute loss of inter-well coherence and $100\%$ build up of diagonal correlations.}
    \label{fig2}
\end{figure}


\section{Measures of order-disorder and complexity} \label{qnt}

Complexity is measured in terms of order and disorder. Often entropy was taken as an appropriate measure of disorder. However, with increase in the number of available states, the disorder of the system increases, as well as entropy increases. Later Landsberg's definition of the disorder parameter ($\Delta$) is well accepted which circumvent the previous problem. 
Disorder is defined as,
\begin{equation}
\Delta = \frac{S}{S_{max}},
\label{disorder}
\end{equation}
where $S$ is the actual information entropy of the system.
$S_{max}$ is the maximum entropy which is accessible to the system. Thus, in the Landsberg definition, entropy and disorder are decoupled. Order is defined as, 
\begin{equation}
\Omega = (1-\Delta).
\label{order}
\end{equation}
$\Omega = 1$ corresponds to perfect ordered and predictable system. $\Omega = 0$ corresponds to complete disorder and randomness. Both order and disorder are size independent and lies between $0$ and $1$. The measure of complexity is further defined appropriately in terms of order and disorder. In the literature, we find three categories of complexity measure as mentioned in the Introduction. To take into account all three categories, we utilize the most generic form of complexity defined as, 
\begin{equation} \label{complexity}
 \Gamma_{\alpha\beta}= \Delta^{\alpha} \Omega^{\beta}= \Delta^{\alpha} (1-\Delta)^{\beta} = (1-\Omega)^{\alpha} \Omega^{\beta}
\end{equation}
It defines complexity of disorder strength $\alpha$ and order strength $\beta$. Thus three categories are subsumed here. With $\beta =0$ and $\alpha >0$, complexity is an increasing function of disorder; with $\alpha =0$ and $\beta >0$, complexity is an increasing function of order. When $\alpha \neq 0$, $\beta\neq 0$, one finds the most common case of `convex' type of complexity. Complexity
\begin{figure}
    \centering
    \includegraphics[scale=0.4, angle=-90]{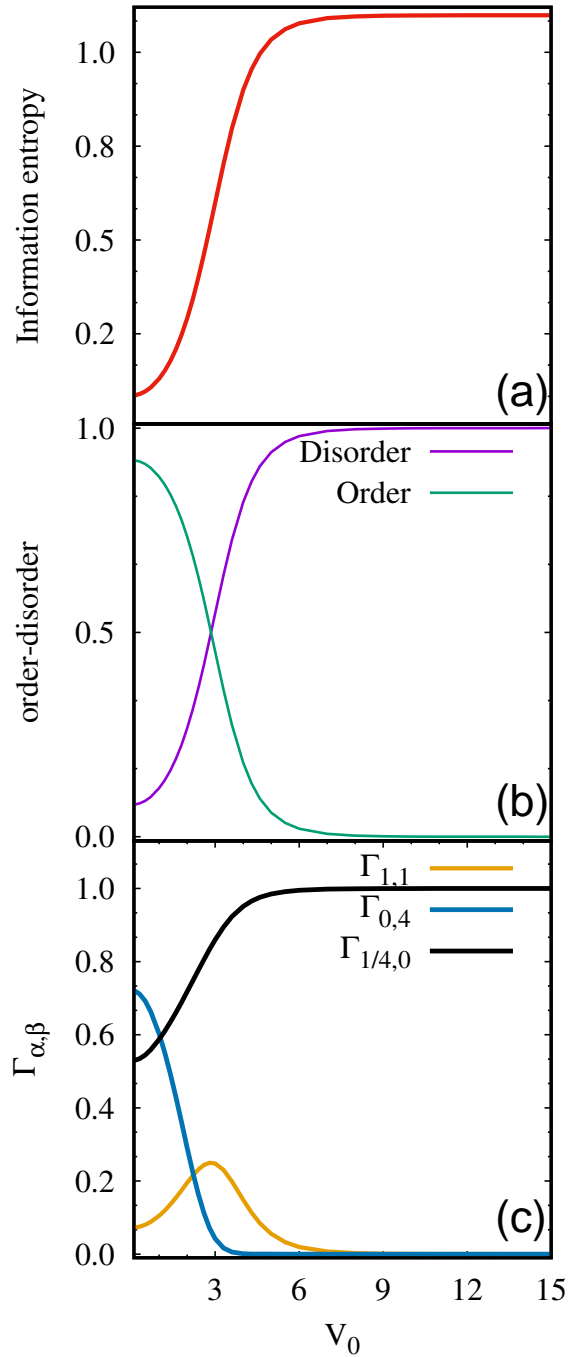}
    \caption{(a) Plot of Shannon information entropy (S) as a function of lattice depth ($V_0$). Entropy is minimum for shallow lattice depth. As lattice depth increases, system transit from SF to MI phase followed by saturation in the entropy at maximum value. (b) Order and disorder of the system of $N=3$ bosons in three well optical lattice. In superfluid to Mott insulator transition, order gradually decreases to zero and disorder gradually increases to one. The intersection point refers as the initiation of SF to MI state. SF phase is characterized as ordered and MI phase is characterized as disordered state. (c) Plot of Complexity measures ($\Gamma_{\alpha \beta}$)  as a function of $V_0$. $\Gamma_{0,4}$, $\Gamma_{1,1}$, $\Gamma_{1/4,0}$ exhibit type I, type II and type III complexity as discussed in the literature~\cite{complexity_PhysRevE}}
    \label{fig3}
\end{figure}
vanishes at zero disorder and zero order and exhibits a maximum in between. However, the key quantity for the calculation of order and disorder are the entropy $S$.  Shannon information entropy both in coordinate space and momentum space are usually taken as the key measure of information entropy. They are defined as $S_x(t) = -\int dx \rho^{(1)}(x,t) ln [\rho^{(1)}(x,t)]$  and  $S_k(t) = -\int dk \rho^{(1)}(k,t) ln [\rho^{(1)}(k,t)]$. Where $\rho^{(1)}(x,t)$ is the one-body density in position space and $\rho^{(1)}(k,t)$ is the same in momentum space. The reduced one-body density in coordinate space  defined as 
\begin{eqnarray}
\begin{split}
\rho^{(1)}(x^{\prime}\vert x;t)=N\int_{}^{}dx_{2}dx_{3}...dx_{N} \psi^{*}(x^{\prime},x_{2},\dots,x_{N};t) \\ \psi(x,x_{2},\dots,x_{N};t).
\label{onebodydensity}
\end{split}
\end{eqnarray}
Its diagonal gives the one-body density $\rho (x,t)$ defined as
\begin{eqnarray}
\begin{split}
\rho( x;t)=N\int_{}^{}dx_{2}dx_{3}...dx_{N}  \psi^{*}(x,x_{2},\dots,x_{N};t) \\ \psi(x,x_{2},\dots,x_{N};t).
\label{onebodydensity2}
\end{split}
\end{eqnarray}
Density distributions are normalized to unity. However, the one-body density are insensetive to address correlations present in the many-body wave function. So we define an alternative measure of many-body information entropy as 
\begin{equation}
 S = - \sum_{i} \bar{n}_i(t) \hspace{0.8ex} ln \hspace{0.2ex} \bar{n}_i(t)
\label{sie}
\end{equation}
This can be called as occupational information entropy as it is defined by the eigenvalues of the reduced one-body density matrix or occupation numbers. For the mean-field theory as there is only one natural occupation, occupational entropy is always zero. In our earlier calculation~\cite{rhombik_jpb}, we have also discussed how time evolution of occupational entropy can be chosen as a good measure for the description of fragmentation. It is also a key quantity in the study of non-equilibrium quench dynamics to establish whether thermalization  and relaxation are ubiquitous in nature. Thus the use of occupational entropy in the measure of order, disorder and complexity is justified.
As pointed out that SF phase exhibits global correlation across the lattice whereas Mott phase exhibits on-site correlations. To explore the link between order-disorder to complexity, we further make an analysis of first order correlation function $g^{(1)}(x_1^{\prime},x_1,t)$ defined as,
\begin{equation}
    g^{(1)}(x_1^{\prime},x_1,t) = \frac{\rho^{(1)}(x_1^{\prime} \vert x_1 ;t)}{\sqrt{\rho(x,t) \rho(x^{\prime},t)}}
\end{equation}
where $\rho$ is the diagonal part of the one-body density matrix given in Eq.~\ref{onebodydensity}. $g^{(1)}(x_1^{\prime},x_1,t)$ quantify how correlated the particles are in the specific system. In recent experiments, it is also possible to calculate higher order correlations experimentally to characterize a many-body system~\cite{hocorr1,hocorr2,hocorr3,hocorr4}.

\begin{figure}
    \centering
    \includegraphics[scale=0.4, angle=-90]{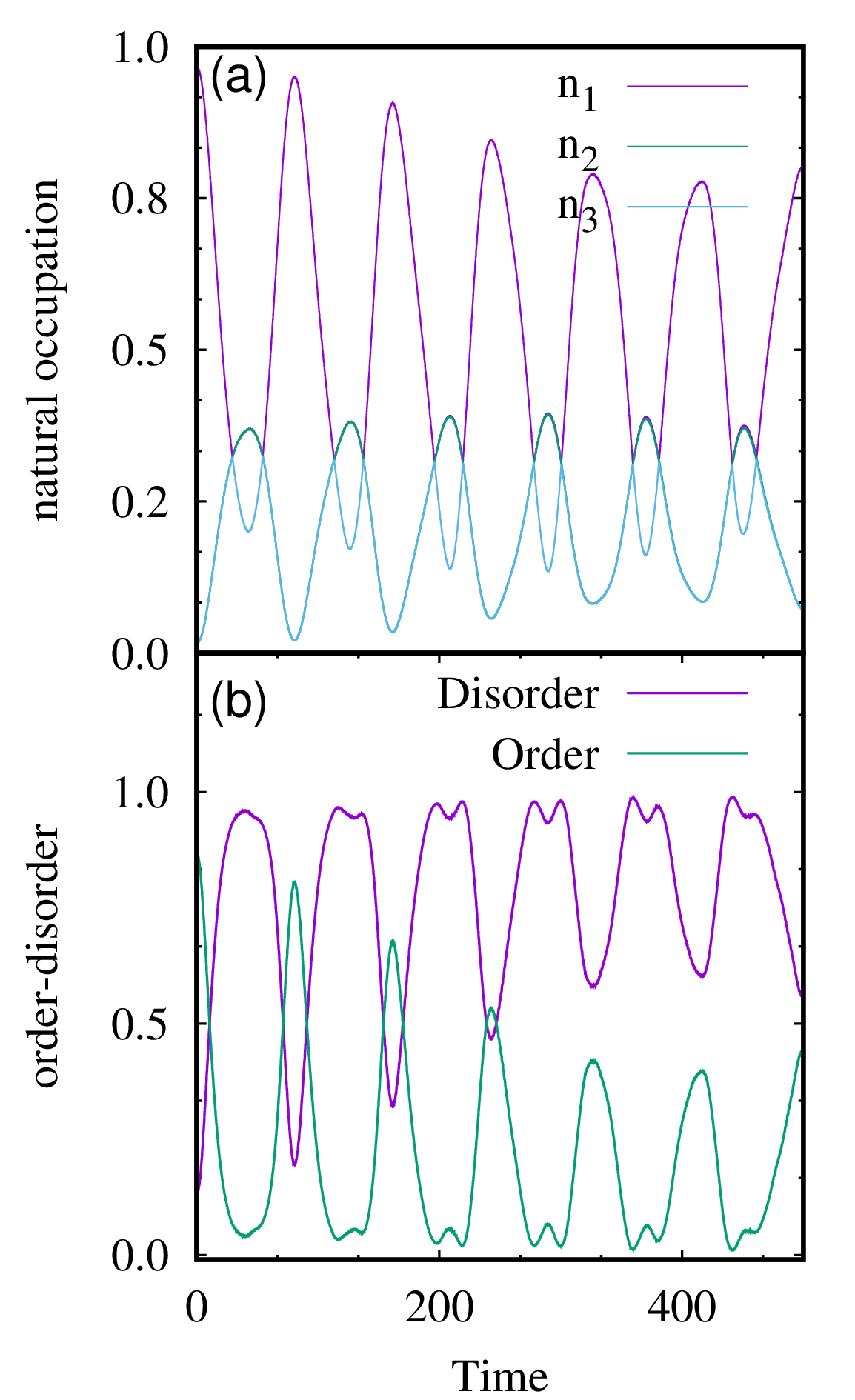}
    \caption{Time evolution of natural occupation, order and disorder of $N=3$ bosons in optical lattice as a function of time. (a) Plot of first three natural orbitals exhibit oscillation. At the crossing point of the first three orbitals, system is in MI phase with $ n_1 \simeq n_2 \simeq n_3 \simeq \frac{1}{3}$. See the text for further discussion. (b) order and disorder plot where order shows maximum value at SF phase and disorder shows maximum value at MI state. The plateau region at the maximum point of the disorder curve tells MI phase holds at that time interval. }
    \label{fig4}
\end{figure}

\section{Results} \label{result}
\subsection{Relaxed state}
The present calculation is performed in the one dimension with $N=3$ repulsively interacting bosons in the optical lattice. The dimensionless strength parameter $\lambda$ of the repulsive interaction is kept fixed to $\lambda = 0.3$ and the lattice depth potential is varied. For the entire manuscript, we consider commensurate filling factor $\frac{N}{W}=1$, i.e., three repulsively interacting bosons in three well. However, the following results are also valid for five repulsively interacting bosons in five wells. As pointed out earlier, fragmentation is the hallmark of MCTDHB, where several natural orbitals exhibit significant population. Thus convergence is an important issue and to capture the correct physics, we need adequate number of orbitals. Before assessing that how much complexity the system is developing during SF to MI transition, we first need to evaluate the many-body information entropy for different lattice depth potential. For the stationary state solution, we follow the improve relaxation method ---- we propagate the MCTDHB equation in imaginary time. For relaxation, we keep $M=6$ orbitals and find that with further increase in number of orbitals, there is no change in the computed quantities.  In Fig.~\ref{fig1}, we plot the occupation of the first, second and third natural orbitals as a function of lattice depth potential. For very small lattice depth, the first orbital has close to $100\%$ occupation and second and third orbitals have insignificant occupation. The many-body wave function can be well approximated by the single orbital mean-field state $\vert N= 3,0,0,0,0,0 \rangle$.
To identify the state, we plot the absolute value of first order correlation function $g^{(1)}(x_1^{\prime},x_1,t)$ for $V_0= 0.7$ in Fig.~\ref{fig2}(a). We observe that coherence within and between the sites is maintained. It confirms the state as superfluid. With increase in lattice depth $V_0$, the occupation in the first natural orbital gradually decreases and contribution from second and third orbitals increase. From the numerical data, we observe that at $V_0=10.0$, the occupation in first three natural orbitals saturates at $33.33\%$. We call that the initial SF state is now three fold fragmented. \\
We further identify that the fragmented many-body state with the configuration $\vert 1,1,1,0,0,0 \rangle$ as Mott state. In Fig.~\ref{fig2}(d), we plot the one-body correlation for $V_0 = 10.0$. It exhibits three separated regimes with non-zero values along the diagonal and zero contribution from the off-diagonal terms. Thus the coherence within the well is maintained and inter-well coherence is completely lost. It is clearly a Mott phase. Thus Fig.~\ref{fig1} and Fig.~\ref{fig2} demonstrate the transition from SF phase to MI phase with gradual change in the lattice depth potential.\\
The main aim of our present work is to characterize the superfluid to Mott transition in terms of order, disorder and complexity. The most important question whether the lattice depth parameter can be chosen as a measure of disorder. In Fig.~\ref{fig3}(a), we plot the many-body Shannon information entropy as a function of lattice depth. For SF phase, entropy is minimum but not zero. With increase in lattice depth, entropy gradually increases and at $V_0=10.0$, it saturates. With further increase in lattice depth parameter, entropy remains at the saturation value. Thus once the MI phase is reached, system retains at that phase even with further increase in lattice depth.\\
In Fig.~\ref{fig3}(b), We plot the order and disorder for varying lattice depth. We observe for SF state, order is maximum and disorder is minimum. As neither the order is one nor the disorder is zero, the SF state is not \textit{perfectly} ordered state. With increase in lattice depth, order gradually decreases and disorder increases. At $V_0=2.8$, order and disorder plot intersects each other which we considered as the initiation of SF to MI state. At $V_0 = 10.0$, order become exactly zero and the disorder becomes exactly $1.0$, which confirm that Mott phase is a \textit{random} phase or a \textit{perfectly} disordered phase. Thus our numerical calculations exhibit that the `SF-MI' transition can be renamed as `order-disorder' transition. The corresponding complexity $\Gamma_{\alpha \beta}$  is plotted in Fig.~\ref{fig3}(c). $\Gamma_{1,1}$ shows convex-type, i.e., type II complexity. For SF phase, complexity is minimum. With increase in lattice depth, complexity increases, reaches a maximum and smoothly reduced to zero for Mott phase. Whereas for $\Gamma_{0,4}$, the complexity with zero disorder exhibits type II complexity of the literature~\cite{complexity_PhysRevE}. $\Gamma_{1/4,0}$, the complexity with zero order, exhibits type I complexity~\cite{complexity_PhysRevE}. Thus we find the existence of all three types of complexity in our calculation.
From the results based on the stationary state calculation, we can conclude that `lattice depth' of the optical lattice can be taken as a measure of disorder in the system. In the language of statistical measure, SF phase is an ordered state and MI phase is a disordered state. From all the related measures in the statistical quantities, we uniquely conclude that SF to MI transition is truly an ordered-disordered transition.
\begin{table*}[ht]
  
    \caption{Superfluid to Mott insulator phase transition time statistics.}
    \label{tab:table1}
    \resizebox{\textwidth}{!}{\begin{tabular}{c|c|c|c|c|c|c}
    \hline
      \textbf{Period} & \textbf{SF to MI initiation} & \textbf{Mott state entry} & \textbf{Mott state exit} & \textbf{MI phase holding time} & \textbf{MI to SF initiation} & \textbf{SF state}\\
      \hline
      First Cycle & 10.6 & 30.0 & 54.0 & 24.0 & 71.0 & 81.0\\
      \hline
      Second Cycle & 91.0 & 113.0 & 137.0 & 24.0 & 154.0 & 162.0\\
      \hline
      Third Cycle & 170.0 & 196.0 & 219.0 & 23.0 & 239.0 & 243.0\\
      \hline
    \end{tabular}}
  
\end{table*}

\begin{figure}
    \centering
    \includegraphics[scale=0.4, angle=-90]{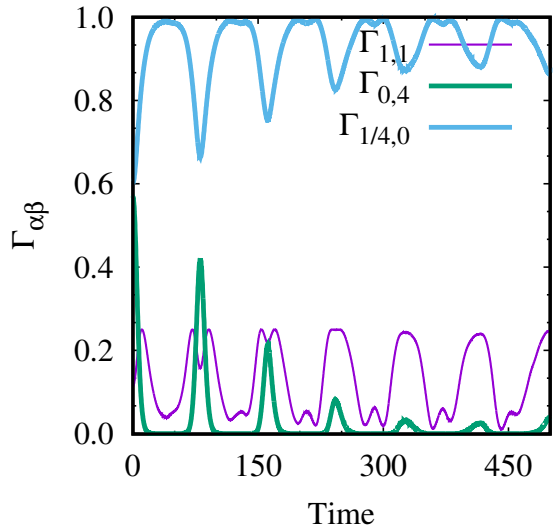}
    \caption{Time evolution of complexity in $\Gamma_{\alpha \beta}$ of $N=3$ bosons in optical lattice as a function of time. $\Gamma_{1,1}$ shows collapse revival dynamics between SF and MI phase. Picks in $\Gamma_{0,4}$ signifies the time when the system is in SF phase. The height of the picks in $\Gamma_{0,4}$ give a qualitative idea about the strength of the SF phase. Picks in $\Gamma_{1/4,0}$ quantify the MI phase.}
    \label{fig5}
\end{figure}

\subsection{Lattice depth quench}
In this section, we do the dynamical measures of order, disorder and complexity. The motivation of the study is to estimate the relative contribution of order and disorder to complexity during evolution. It will practically demonstrate how the system would optimize the complexity which basically will exhibit the ability or inability of the system to self organize with the external change. In the quench dynamics, we monitor the time evolution of the natural orbitals and entropy production. In the dynamical evolution, the convergence is a serious issue and we need $M=12$ orbitals for quench dynamics. We report long-time dynamics (up to t=500) in all measured quantities. We initially prepare the system in SF phase with weak interaction strength $\lambda= 0.1$ in shallow lattice depth of $V_0 = 3.0$. We quench the system to MI phase by sudden increase in lattice depth to $V_0= 10.0$ at $t = 0$. In Fig.~\ref{fig4}(a), we plot the natural occupation in first three orbitals as a function of time. At $t=0.0$, only the first natural orbital contributes, which corresponds to SF phase. With increase in time, fragmentation occurs. We observe that at $t=30$, the system is a fully fragmented MI state, the lowest three natural orbitals $n_1 , n_2, n_3$ have close to $33\%$ population. Between $t=30$ to $t=54$, $n_1, n_2$ overlap, $n_3$ becomes down. From the one-body correlation plot (not shown here), between  $t=30$ and $t=54$, we find that the system exhibits only the diagonal correlation, which signifies the system is in MI phase. So, for this choice of parameter, MI state is retained in this interval. Then at $t=81$, it enters to SF phase with maximum occupation only in the first orbital, at $t=113$ it again enters the second MI phase and retained the MI phase till $t=137$. The above scenario repeats with entry and exit in SF and MI phase. This basically simulates collapse revival dynamics as observed in Greiner's experiment~\cite{greiner2}. The time evolution of the many-body information entropy exhibits  the same collapse revival picture in the same time scale (not shown here).\\
From the time evaluation of many-body information entropy, we further calculate the  time dependence in order, disorder and complexity. Qualitatively, all the measures exhibit collapse revival scenario. However the observations based on the time scale need further explanation. From Fig.~\ref{fig4} (b), initially when the many-body state is in SF phase, only one natural orbital contributes, order is maximum and disorder is minimum. With increase in time, order decreases and disorder in the system builds up. At $t=10.6$, order-disorder cross each other. From the natural occupation, we find the significance of the point of crossing.
$t=10.6$ is the initiation of Mott phase when fragmentation starts ----
order-disorder exhibit the same weights ($\simeq 50\%$). Further at $t=30$, disorder becomes close to one and order is close to zero which signify the pure Mott phase. Between $t=30$ and $t=54$, both the order and disorder have a plateau region which signify the holding time of Mott phase. Then at $t=71$, order-disorder again cross each other which is the initiation of MI to SF phase. 
At $t=81$, the system enters in SF phase exhibiting maximum order and minimum disorder. Then again order decreases and disorder increases further. At $t=113$, system enters in second MI phase. The time interval $t=113$ to $t=137$ is the holding time of second MI phase. 
The time scale for different phases during the time evolution is presented in Table.\ref{tab:table1} for the first three cycles. All the measured time scales are in mutual agreement. 
Although the time dynamics exhibit the collapse revival picture on an average, however close scrutiny reveals some additional information. In the time evolution of order, the peak which corresponds to SF phase gradually decreases. Initially the peak value was close to one, finally it settle down to 0.4.  It exhibits that the system initially has full ability to turn back to SF phase, however with complex time evolution, it looses its ability gradually. We understand that in the long-time dynamics, many excited states contribute in a complex manner. So it is very hard for the system to retain its perfect order and coherence. The study of the evolution of the natural occupation and entropy evolution was not so sensitive tool to detect the above observation.\\
Fig.~\ref{fig5} shows the time evolution of complexity ($\Gamma_{1,1}$, $\Gamma_{0,4}$ and $\Gamma_{1/4,0}$). $\Gamma_{1,1}$ repeats the common-type (type II) behaviour in the dynamics and exhibits same features in same time scale as observed in the time evolution of order and disorder. We do not find any steep building in complexity. By sudden increase in lattice depth means we are pumping energy to the system externally and it will be distributed through one-body term in the Hamiltonian. 
The time dynamics of $\Gamma_{1,1}$ exhibits that the system is able to distribute the extra energy between the interacting bosons and thus able to self organize the external perturbation.\\ 
For further investigation we calculate complexity $\Gamma_{0,4}$ of disorder weight factor zero and order of weight factor four to quantify the absolute contribution by order only.  Peaks in $\Gamma_{0,4}$ tell about the strength of order present in the system. The first peak in $\Gamma_{0,4}$ arises at $t=81$, which is SF phase. Second and third peak positions are at $t=162$ and $t=343$ respectively. All the peaks indicates the formation of SF phase which is an ordered state. In the long time dynamics, $\Gamma_{0,4}$ dies out to zero gradually which indicates the gradual disappearance of SF phase. It also indicates the gradual loss of coherence in the time evolution.\\
We next calculate $\Gamma_{1/4,0}$, which defines complexity with  order factor of zero and disorder factor of small finite quantity of 0.25. It will facilitate to quantify the contribution of disorder to complexity. Initially dynamics of $\Gamma_{1/4,0}$ is reverse to that of dynamics in $\Gamma_{0,4}$ qualitatively. The peaks correspond to Mott state with maximum disorder. The peaks are flat as the Mott state is hold for some time (as discussed earlier) before switching to SF phase. In the long time dynamics it gradually settle to one.\\
Thus we conclude that the above analysis nicely exhibit how the system tends to self organize on the sudden quench in optical lattice. Although initially all the complexity measure point out the same physics, collapse and revival dynamics, however in the long time, the system will gradually loose its ability of revival and finally will settle to MI phase only. 
\section{Conclusion} \label{conc}
`Statistical complexity' is one of the most circulating word in scientific research of Physics, biology, mathematics, computer science etc. Although there is no strictly followed definition of `what is complexity', it is defined in many ways in the literature. From its vast application, it is found that SDL measure of complexity is well understood.

SDL measure has been extensively applied in different systems including atoms and molecules. Here we consider ultracold trapped atoms in the optical lattice which has been proved as a most challenging platform to study the many-body physics. Interacting bosons in optical lattice exhibit different quantum phases like superfluid and Mott insulator phase. The quantum phase transition has been experimentally studied as well as there are numerous theoretical calculation on the dynamical evolution. Although most of the calculation are based on mean-field level or utilizing Bose Hubbard model, the strongly interacting bosons in shallow lattice deserve quantum many-body calculation. We report our results on small ensemble of few particle system utilizing MCTDHB method which is exact by its construction and retain many-body correlation. We obtain many-body states which are the few-body analogy of different thermodynamic phases.\\
The main motivation of our work is how to consult the various quantum phases with the measure of order-disorder and complexity. As the lattice depth is an easily controllable parameter experimentally, our theoretical analysis both for the relaxed state as well as dynamically evolve state exhibit that lattice depth can act as order-disorder parameter. SF phase is characterized as an ordered but not properly ordered state, whereas Mott phase is characterized as proper disordered state. Complexity exhibits type II features. SF to MI phase transition can be renamed as `order-disorder' transition. In the quench dynamics, we study the entire physics behind the SF to MI transition in the light of dynamical measures of order, disorder. We find that initially the system is able to demonstrate the collapse to Mott phase and revival to superfluid phase. However, the system gradually looses its ability and finally looses global correlation across the lattice and settles to Mott insulator phase in long-time dynamics. We also able to present the time scale of different phases through several cycles.

\begin{acknowledgements}
R. Roy acknowledges the University Grant Commission (UGC) India for the financial support as a senior research fellow. 
\end{acknowledgements}
\bibliography{manuscript}
\bibliographystyle{spphys}
\end{document}